\documentstyle[aps,floats,epsf,twocolumn]{revtex}

\begin{document} \draft

\title{Phase Fluctuations and Spectral Properties of Underdoped Cuprates}

\author{M. Franz and A. J. Millis}
\address{Department of Physics and Astronomy, Johns Hopkins University,
Baltimore, MD 21218
\\ {\rm(\today)}
}
%
\address{~
\parbox{14cm}{\rm
\medskip
We consider the effect of classical phase fluctuations 
on the quasiparticle spectra of underdoped high-$T_c$
cuprate superconductors in the pseudogap regime above $T_c$. We show that 
photoemission and tunneling spectroscopy data are well accounted for by a 
simple model in which mean field $d$-wave quasiparticles are semiclasically
coupled to supercurrents induced by fluctuating
unbound vortex-antivortex pairs. We argue that the data imply that transverse 
phase fluctuations are important at temperatures above $T_c$, while 
longitudinal fluctuations are unimportant at all temperatures.  
}}
\maketitle


%
\narrowtext
\section{Introduction.}
It is now a well established experimental fact that the underdoped cuprate 
superconductors exhibit a ``pseudogap'' behavior above the superconducting 
critical temperature $T_c$, which is  characterized by
a vanishing superfluid density as obtained by transport measurements, 
but persistence of a gap in the quasiparticle excitation spectrum
as measured by various spectroscopies\cite{randeria1}. Recent angle resolved 
photoemission (ARPES)
\cite{ding1,norman2,zx1,norman1} and scanning tunneling spectroscopy (STS)
\cite{renner1,miyakawa1}
experiments indicate that in the underdoped cuprates 
the gap evolves smoothly as the temperature is increased through $T_c$.
Although the gap begins to fill in and the sharp quasiparticle peaks are lost
above $T_c$, the position of the gap edge changes only slightly. 
This is in sharp contrast to the behavior of 
overdoped cuprates and conventional
superconductors where the gap {\em closes} at $T=T_c$.
It is further found that the gap {\em increases} as the doping concentration
is reduced from its optimum value, while at the same time  $T_c$ decreases.
This results in highly anomalous ratios $2\Delta/k_BT_c$ which were reported 
to attain values of 12 or more in Bi$_2$Sr$_2$CaCu$_2$O$_{8+\delta}$
(BiSCCO), compared 
to the weak coupling BCS value of 3.54. ARPES results also indicate that the
angular dependence of the gap function on the Fermi surface,
which below $T_c$ follows the  $\Delta_{\bf k}=\Delta_d\cos(2\theta)$ 
shape expected for 
a $d_{x^2-y^2}$ order parameter, develops extended gapless regions around
the nodes above $T_c$ whose size increases with $T$. Although a 
 number of important 
experimental issues remains to be settled, such as the temperature at which 
the pseudogap closes and its possible persistence in optimally and even
overdoped cuprates suggested by the recent STS results\cite{renner1}, the
basic picture of a superconducting quasiparticle gap persisting over a 
wide range of $T>T_c$ in underdoped materials is well established. 

Many theoretical concepts, including spin 
fluctuations\cite{chubukov1}, condensation
of preformed pairs\cite{geshkenbein1}, SO(5) symmetry\cite{zhang1}
and spin-charge separation\cite{lee1}, have been invoked to explain the 
pseudogap behavior.
In this paper we study the implications of a scenario put forward by Emery and 
Kivelson\cite{emery} who, following earlier work of Uemura and 
co-workers\cite{uemura}, proposed  that 
 the underdoped material above $T_c$ is in a state with a non-zero
{\em local} amplitude of superconducting pairing, but is not
truly superconducting due to thermal fluctuations in the phase of the
order parameter.
Within such a scenario the transition at $T_c$ is of the 
Kosterlitz-Thouless (KT) type, slightly rounded by the weak coupling
between the copper-oxygen planes along the $c$-axis. In a strictly 2D system
the KT transition is associated with proliferation of unbound
vortex-antivortex pairs. Weak coupling between the planes leads to correlated
motion between vortices in adjacent planes which form
3D vortex loops close to the critical temperature. The transition
to the disordered phase is then characterized by the 
appearance of vortex loops with arbitrarily large radii\cite{blowup}.

In the present paper we study the spectral properties of a superconductor
in the incoherent state above the phase disordering transition but below
the mean field transition at which the local gap forms.
We find that fluctuating currents arising from unbound vortex-antivortex pairs
can contribute significantly to the ARPES and STS lineshape broadening in the 
low-energy region of the spectrum.
By analyzing the experimental data we estimate the strength of these 
superconducting phase fluctuations and we deduce the vortex core 
energy.

We model the underdoped cuprate superconductor as a set of independent 
2D superconducting layers, each undergoing a KT transition at a temperature
$T_{\rm KT}$ which we identify with the superconducting critical temperature
$T_c$. The weak interplane coupling, which we neglect, will
affect the very long lengthscale physics (changing e.g. the universality
class of the transition from KT to 3D XY), but should not affect the
shorter lengthscale fluctuations which contribute to the electron spectral 
functions of interest here. The disordered state above
$T_{\rm KT}$ can be thought of as a ``soup'' of fluctuating vortices with
positive and negative topological charges and with total vorticity constrained
to zero. Each of these vortices is
surrounded by a circulating supercurrent which decays as $1/r$ with the 
distance from the core. Such supercurrents, within a semiclassical
approximation, lead to a Doppler-shifted local quasiparticle 
excitation spectrum of the form\cite{degennes,tinkham} 
\begin{equation}
E_{\bf k}=E_{\bf k}^0+\hbar{\bf k}\cdot{\bf v}_s({\bf r}),
\label{en}
\end{equation}
where ${\bf v}_s({\bf r})$ is the local superfluid velocity and $E_{\bf k}^0
=\sqrt{\epsilon_{\bf k}^2+|\Delta_{\bf k}|^2}$ is the usual BCS spectrum.
The change in the local excitation spectrum will affect the spectral
properties of the superconductor in that the physically relevant spectral 
function must be averaged over the positions of fluctuating vortices. 
 
This effect will be particularly pronounced 
in a $d$-wave superconductor since  Eq.\ (\ref{en})
implies formation of a region on the Fermi surface with
$E_{\bf k}<0$ around a nodal point for arbitrarily small ${\bf v}_s({\bf r})$. 
Physically this corresponds to a region of gapless excitations on 
the Fermi surface which leads to a finite  density of states (DOS)
at the Fermi level.  
As first discussed by Volovik\cite{volovik1}, a similar situation arises 
in the {\em mixed} state of a $d$-wave superconductor 
where the superflow around the field-induced vortices 
leads to the residual DOS proportional to $\sqrt{H}$.
This unusual field dependence arises because the distance between vortices
in the vortex lattice $d_v\sim H^{-1/2}$ and 
the average superfluid velocity projected onto a gap node 
direction is proportional to $d_v^{-1}$.
At low $T$ and high field this implies a 
$\sim T\sqrt{H}$ contribution to the electronic specific heat which was 
indeed observed in the measurements on YBa$_2$Cu$_3$O$_{6.95}$ 
single crystals\cite{moler,revaz}. 
In the present case, instead of a regular Abrikosov 
lattice of field-induced vortices, we consider a fluctuating plasma of 
thermally induced vortices and antivortices. The essential physics however 
remains the same.

\section{Theory: quasiparticle excitations coupled to phase fluctuations.} 
We shall be interested in how the supercurrents induced by
phase fluctuations affect the spectral function of a superconductor,
$A({\bf k},\omega)=-\pi^{-1}{\rm Im}{\cal G}({\bf k},\omega)$,
which may be measured by ARPES and STS experiments.  Here 
${\cal G}({\bf k},\omega)$ is the diagonal part of the full 
superconducting Green's function which solves the  
Gorkov equations for a $d$-wave superconductor, given in the Appendix. 
In the mean field approximation (neglecting, among other things, phase
fluctuations) the diagonal Green's function may be written as
\begin{equation}
{\cal G}_0^{-1}({\bf k},\omega)= \omega-\epsilon_{\bf k}+i\Gamma_1-{\Delta_{\bf k}^2\over
\omega+\epsilon_{\bf k}},
\label{g0}
\end{equation}
where, following \cite{norman1}, we have added to the usual mean-field
solution a single particle scattering rate $\Gamma_1$.
Note that this form of the scattering rate in ${\cal G}_0$ constitutes
an non-trivial assumption. It is not pairbreaking, in the sense that it is
ineffective at small $\omega$, $\epsilon_{\bf k}$; i.e. in the region 
$\omega< E_{\bf k}\sim\sqrt{\epsilon_{\bf k}^2+\Delta_{\bf k}^2}$. By 
contrast in a 
$d$-wave superconductor, a conventional scattering rate enters via the 
replacement $\omega\to\omega+i\Gamma$, leading to a broadening which is 
effective even at low $\omega$, $\epsilon_{\bf k}$. 
As shown by Norman {\em et al.}\cite{norman1} the form given in Eq.\ 
(\ref{g0}) agrees with the ARPES data at $T<T_c$.  
We demonstrate below that it also agrees with STS. 

At $T>T_c$ Norman {\em et al.}\cite{norman1} showed that additional 
{\em pairbreaking} scattering is needed to account for the ARPES data, 
which they modeled phenomenologically by introducing another scattering rate 
$\Gamma_0\ne\Gamma_1$, making a 
replacement $\omega\to\omega+i\Gamma_0$ in the last term of Eq.\ (\ref{g0}).
They suggested that $\Gamma_0$ could arise from exchange of pair fluctuations;
we find, by explicitly evaluating the corresponding propagator\cite{millis2},
that this proposed mechanism does not 
account for the observed magnitude of $\Gamma_0$. This conclusion is supported
by the results of Vilk and Tremblay\cite{vilk1}.  

We now discuss what we believe to be a more likely source of the pairbreaking 
scattering, namely supercurrents induced by phase fluctuations.  
In order to determine how ${\cal G}_0$ is changed in the presence of 
superflow it is useful to recall the origin of the energy shift in
Eq.(\ref{en}). This can be  derived\cite{degennes,tinkham}
by assuming a state of {\em uniform} superflow with ${\bf v}_s=\hbar{\bf q}/m$
\cite{note4}
induced by an order parameter of the form $e^{2i{\bf q}\cdot{\bf r}}
\Delta_{\bf k}$.
By solving the appropriate set of Bogoliubov-de Gennes equations and 
retaining only terms to linear order in ${\bf q}$, one finds that the energy 
is modified as indicated in (\ref{en}) while the coherence factors are 
to the same order unchanged. This result is then semiclassically
extended to non-uniform situations by assuming slow spatial variations of 
${\bf v}_s({\bf r})$. 

One can follow this exact procedure and solve the appropriate 
Gorkov equations for ${\cal G}_{\bf q}$ in the presence of superflow. 
One finds (see Appendix) the
following intuitively plausible result which is exact for
uniform flow up to terms linear in ${\bf q}$\cite{note1}:
\begin{equation}
{\cal G}_{\bf q}({\bf k},\omega)={\cal G}_0({\bf k}-{\bf q},\omega-\eta),
\label{gq}
\end{equation}
where $\eta\equiv\hbar{\bf v}_F({\bf k})\cdot{\bf q}\simeq\hbar{\bf k}\cdot
{\bf v}_s$. Here
${\bf v}_F({\bf k})=(\partial\epsilon_{\bf k}/\partial{\bf k})_{k=k_F}\simeq
\hbar{\bf k}_F/m$ 
is the Fermi velocity and the last equality holds when the Fermi surface is
approximately isotropic.   In the following
we shall assume that Eq.\ (\ref{gq}) can be applied locally when 
${\bf v}_s({\bf r})$ varies slowly in space. 
Applying the above prescription to (\ref{g0}) one finds, again to the leading 
order in ${\bf q}$, 
\begin{equation}
{\cal G}_{\bf q}^{-1}({\bf k},\omega)= \omega-\epsilon_{\bf k}+i\Gamma_1-
{(\Delta_{\bf k}-\zeta)^2\over
\omega+\epsilon_{\bf k}-2\eta},
\label{gqq}
\end{equation}
where $\zeta\equiv{\bf v}_\Delta({\bf k})\cdot{\bf q}$ with 
${\bf v}_\Delta({\bf k})=(\partial\Delta_{\bf k}/\partial{\bf k})_{k=k_F}$. 
One can easily estimate
$v_\Delta/v_F\sim (\xi_0k_F)^{-1}\sim\Delta_d/\epsilon_F$ which is typically
a small number in superconductor. We therefore expect that $\zeta\ll\eta$.
A more detailed numerical analysis indeed shows that, as long as
$\Delta_d/\epsilon_F$ is small compared to unity, the effect of $\zeta$ on
the spectral lineshape is negligible compared to that of $\eta$, and will 
be dropped in the following. 

A typical experimentally measured quantity, such as the ARPES or STS lineshape,
will provide information on ${\cal G}_{\bf q}$ {\em averaged} over the phase
fluctuations. Thus, we need to evaluate 
\begin{equation}
\bar{\cal G}_{\bf q}({\bf k},\omega)
=\int d\eta P(\eta){\cal G}_{\bf q}({\bf k},\omega),
\label{avg}
\end{equation}
where $P$ is the probability distribution of $\eta$ given by
\begin{eqnarray}
P(\eta)&=&\langle \delta[\eta-\hbar{\bf k}\cdot{\bf v}_s({\bf r})]\rangle.
\label{p2}
\end{eqnarray}
The angular brackets indicate
 thermodynamic averaging over the phase fluctuations in the ensemble 
specified by the 2D XY Hamiltonian\cite{emery}
\begin{equation}
{{\cal H}_{XY}\over k_BT}={1\over2}V\left(2m\over \hbar\right)^2
\int {\bf v}_s^2({\bf r}) d^2r,
\label{ham}
\end{equation}
where ${\bf v}_s({\bf r})$ is understood to contain both longitudinal 
(spin wave like)
and transverse (vortex like) excitations. $V=V_0/k_BT$ is a dimensionless
coupling constant and $V_0$ is related to the superfluid density $n_s$ by
$V_0=\hbar^2n_s/4m$. In the nearest neighbor XY model, 
$k_BT_{KT}\simeq 0.9V_0$\cite{emery}; more generally  
$k_BT_{KT}=(\pi/2)V_0(T=T_{\rm KT})$\cite{minnhagen1}. Longitudinal phase
fluctuations result in the spatial modulation of charge density and
will be therefore suppressed by Coulomb interaction at long wavelengths. This
interaction is not explicitly included in the XY Hamiltonian (\ref{ham}) but
we return to it shortly.

The last term in Eq.\ (\ref{g0}) can be thought of as a superconducting 
self energy $\Sigma_s(\omega,{\bf k})$. Eqs.\ (\ref{gqq},\ref{avg}) then imply 
that the primary
effect of the phase fluctuations is to smear the functional dependence of 
$\Sigma_s(\omega,{\bf k})$ on the energy variable, broadening 
the spectral lineshape. A more detailed analysis shows that $\eta$ acts 
primarily to {\em fill in}
 the gap, in a way similar to the inverse pair lifetime 
$\Gamma_0$ introduced by phenomenological considerations 
in Ref.\cite{norman1}. $\Gamma_1$, on the other hand,
does not affect the lineshape at low energies: notice that 
${\cal G}_0({\bf k}_F,\omega=0)=0$ for any $\Gamma_1$. 

We now
give a quantitative description of this broadening by explicitly evaluating
$P(\eta)$ and the resulting lineshapes as a function of temperature. 
Making use of the identity $\delta(x)=(2\pi)^{-1}\int ds e^{isx}$ Eq.(\ref{p2})
becomes
\begin{equation}
P(\eta)={1\over 2\pi}\int_{-\infty}^{\infty}ds e^{is\eta}\left\langle 
e^{-is\hbar{\bf k}\cdot{\bf v}_s({\bf r})}\right\rangle.
\label{p1}
\end{equation}
To the leading order in cumulant expansion one can write 
$\langle e^{-is\hbar{\bf k}\cdot{\bf v}_s({\bf r})}\rangle=
\exp[-{1\over 2}s^2\hbar^2
k_\alpha k_\beta\langle v_s^\alpha v_s^\beta\rangle]$, where summation over 
repeated indices is implied and the spatial variable has been suppressed.  
This statement becomes exact when the transverse fluctuations can be 
represented by Gaussian degrees of freedom, as is done in the 
Debye-H\"{u}ckel approximation employed below. 
The $s$-integral in (\ref{p1}) can now be explicitly carried out,
yielding a Gaussian distribution
\begin{equation}
P(\eta)=(2\pi W)^{-1/2}e^{-\eta^2/2W},
\label{p11}
\end{equation}
with $W=\hbar^2 k_\alpha k_\beta\langle v_s^\alpha v_s^\beta\rangle$. 
We have thus
reduced the problem of finding the probability distribution $P$ to 
evaluation of a correlator $\langle v_s^\alpha v_s^\beta\rangle$. 
For a 2D system described by the Hamiltonian
(\ref{ham}), this correlator
has been considered by Halperin and Nelson \cite{halperin1}. They found,
using a Debye-H\"{u}ckel approximation valid for  for $T$ well above $T_{KT}$,
\begin{equation}
\langle v_s^\alpha({\bf p}) v_s^\beta(-{\bf p})\rangle=
{\hbar^2\over 2m^2V}
\left[{p_\alpha p_\beta\over p^2} +{\delta_{\alpha\beta}
-p_\alpha p_\beta/ p^2\over 1+c_3\xi_c^2p^2/4\pi^2V}
\right],
\label{cor1}
\end{equation}
where ${\bf v}_s({\bf p})=\int d{\bf r} e^{-i{\bf p}\cdot{\bf r}}
{\bf v}_s({\bf r})$.
The first term in brackets comes from the longitudinal and the second from 
the transverse fluctuations.   $c_3=(2E_c-\pi^2V_0)/k_BT$
is a dimensionless quantity related to the density of vortices, 
$E_c$ is the vortex core energy and  $\xi_c$ is the core cutoff.

Upon averaging over fluctuations the 
translational invariance is restored and we may evaluate the real 
space correlator at ${\bf r}=0$:
\begin{equation}
\langle v_s^\alpha(0) v_s^\beta(0)\rangle=
{1\over(2\pi)^2}\int d^2p\langle v_s^\alpha({\bf p}) v_s^\beta(-{\bf p})
\rangle.
\label{cor2}
\end{equation}
Explicit integration finally yields, for $|{\bf k}|=k_F$,
\begin{equation}
W={\pi^3\Delta_d^2\over8V}\left[1+{V\over c_3}\ln\left(1+{c_3\over V}
\right)\right].
\label{W}
\end{equation}
The short wavelength divergence in (\ref{cor2}) has been cut off at 
$q_c=2\pi/\xi_c$ and we used the BCS relation $\xi_c=\hbar v_F/\pi\Delta_d$. 
The first term in brackets comes from longitudinal fluctuations and would
be suppressed in realistic models in which the Coulomb interaction is
important. 
The second term, $\alpha_v\equiv(V/c_3)\ln(1+c_3/V)$, comes from transverse
fluctuations due to vortices, and is always positive and smaller then
1 (this follows since $c_3>0$ is required by the stability of the system). 
To the extent that $E_c$ is independent of temperature the ratio $c_3/V$, and 
therefore $\alpha_v$, is 
$T$-independent. The primary $T$-dependence of $W$ therefore comes from 
$V$ in the prefactor.  After expressing $V_0$ in terms of $T_{\rm KT}$, $W$
may be written as
\begin{equation}
W\simeq 3.48(1+\alpha_v)(T/T_{KT})\Delta_d^2. 
\label{W1}
\end{equation}

Eqs.\ (\ref{gqq}), (\ref{avg}) and (\ref{p11}) describe the effect of 
classical 
phase fluctuations on the spectral function of a superconductor. From the
knowledge of such a spectral function one can compute the respective ARPES
and STS lineshapes, extract the parameter $W$, and compare it with the 
prediction given by Eq.\ (\ref{W1}). This will be done in the next section,
but we first discuss the validity and some qualitative aspects of the results 
presented above. 

Our results depend 
on three assumptions; that the electron Greens function may be 
calculated using semiclassical methods, that the phase fluctuations are
quasi-static, and that there is a sufficiently wide KT temperature regime 
in which reasonably well defined vortex-antivortex plasma exists. The first of
these assumptions applies when the coherence length is sufficiently larger
than the inter electron spacing (i.e. $k_F\xi\gg 1$) and is the same assumption
as underlies Volovik's prediction\cite{volovik1}
 of a $\sim T\sqrt{H}$ dependence of the
specific heat in the mixed state of a $d$-wave superconductor. As this behavior
is observed in YB$_2$Cu$_3$O$_7$\cite{moler,revaz}, 
we believe that this assumption is 
well justified. The second assumption, of quasi-static phase fluctuations, has
two parts. The transverse fluctuations come from vortices, so for them the
essential assumption is that vortices move slowly compared to electrons. This
is justified by Bardeen-Stephen results, which imply that vortex motion is
overdamped and thus diffusive, and so surely slower than the ballistic motion
of the electrons. For longitudinal fluctuations the situation is less clear. 
If the hypothesis of Emery and Kivelson, that they are classical (i.e.
that at $q\sim \xi_c^{-1}$ we have $\omega_q\ll T$)\cite{emery} is accepted
then Eq.\ (\ref{cor1}) 
applies. However, as we shall see, the data contradict this. 
A more likely scenario is that Coulomb interaction pushes the longitudinal 
fluctuations up to the plasma frequency, in which case the coupling to 
electrons is very weak and one should simply remove the longitudinal 
fluctuations from the theory. We shall see that the data are consistent with 
this picture. If (for some as yet unknown reason) the longitudinal fluctuations
are collective modes with a velocity of the order of the Fermi velocity, then
our results do not apply.
The final assumption, of a wide temperature regime between the mean-field
and KT transitions, is the most difficult to justify, except on empirical
grounds. This hypothesis was proposed in Ref.\cite{emery} and our results
are consistent with it. The theoretical justification for a clean system
must involve proximity to a Mott insulating phase, which suppresses the
superfluid stiffness and hence $T_{KT}$, but does not suppress the pairing.
A detailed theoretical treatment in two dimensions has not been given. 
 
As mentioned above, $W$ given by Eq.\ (\ref{W1}) describes both longitudinal
and transverse fluctuations of the phase and can be written accordingly as
$W=W_L+W_T$, where
\begin{eqnarray}
W_L&\simeq & 3.48(T/T_{KT})\Delta_d^2,\label{W2}\\
W_T&\simeq & 3.48\alpha_v(T/T_{KT})\Delta_d^2,\label{W3}.
\end{eqnarray}
The expression for $W_T$ is valid at temperatures well above $T_{\rm KT}$,
where all pairs can be thought of as unbound and thermal energy dominates
over the intervortex interaction\cite{minnhagen1}. 
Well below $T_{\rm KT}$, on the other hand,
we expect $W_T=0$ since the vortices appear only in tightly bound pairs
which contribute negligible supercurrent beyond the lengthscale set  by the
pair size  and the pair density is exponentially small 
$\sim e^{-2E_c/k_BT}$. By numerically integrating the appropriate scaling 
relations\cite{halperin1} one could in fact obtain $W_T$ at all 
temperatures. However such level of detail is beyond the scope of this paper
and we shall confine ourselves to the limiting cases stated above and note that
$W_T$ has nonsingular monotonic behavior across $T_{\rm KT}$. 

The expression
(\ref{W2}) for $W_L$ is expected to be valid down to low temperatures, 
provided quantum fluctuations and the Coulomb interaction
can be neglected. In a $d$-wave
superconductor the temperature dependence of $W_L$ will be modified by the
$T$-linear temperature dependence of the superfluid density\cite{bonn} 
$n_s\sim \lambda^{-2}(T)$ which enters the definition of $V_0$ in the
XY Hamiltonian (\ref{ham}). It is interesting to note that at 
temperatures below $T_c$, say at $T=T_c/2$, Eq.\ (\ref{W2}) implies
$W_L\simeq\Delta_d^2$, i.e. large broadening of the spectral function by
longitudinal fluctuations. Such a large broadening, comparable to the 
gap itself, would completely 
obliterate any signature of the gap in the excitation spectrum. Clearly,
this is not observed experimentally\cite{ding1,norman2,renner1}. As 
shown below and in Ref.\ \cite{norman1}, experimental data are consistent 
with $W=0$ below $T_c$. We must therefore 
conclude that longitudinal fluctuations are strongly suppressed by the
Coulomb interaction as suggested in Ref.\cite{millis1}. On the same grounds
we may argue that the observed linear 
temperature dependence of the magnetic penetration 
depth\cite{bonn} is not due to phase fluctuations, as suggested by some authors
\cite{roddick1,emery2}, but due to thermally excited quasiparticles
in the nodes of the $d$-wave gap\cite{hirschfeld}. 
Transverse fluctuations, on the other hand, result in
no net charge displacement and are thus unaffected by Coulomb interaction.

\section{Comparison to experimental data}
\subsection{Photoemission}

As shown by Norman and co-workers\cite{norman2}, a particularly simple 
relationship exists between the spectral function of a superconductor and a 
{\em symmetrized} ARPES spectrum at the Fermi surface,
\begin{equation}
 I_S(\omega)\propto{\bar A}({\bf k}_F,\omega),
\label{sym}
\end{equation}
where $I_S(\omega)=I(\omega)+I(-\omega)$ and $I(\omega)$ is the measured
ARPES lineshape. The advantage of this symmetrized representation is that
the thermal broadening due to the Fermi functions is automatically subtracted
out. The proportionality (\ref{sym}) is expected to hold for $|\omega|$
less then few tens of meV and remains valid in the presence of a symmetric
energy resolution function. We first discuss qualitative features of our
predicted lineshape and the experimental data, and then present results
of a detailed fitting procedure. 

\begin{figure}[t]
\epsfxsize=9.5cm
\epsffile{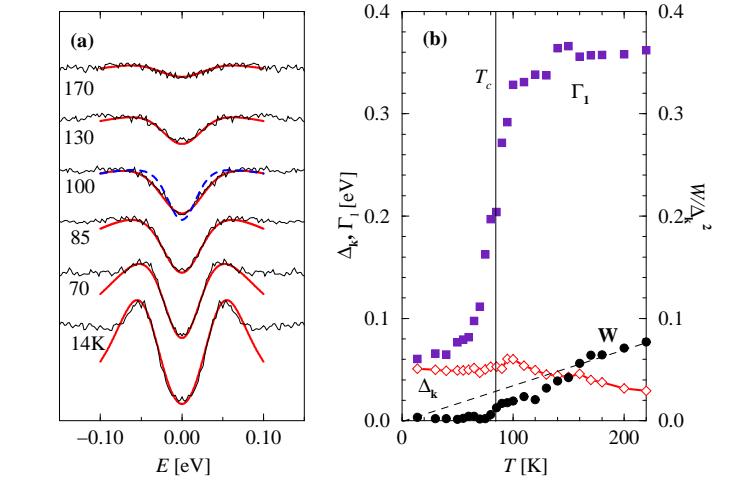}
\caption[]{a) Symmetrized ARPES lineshapes $I_S(\omega)$ 
at momentum near $(\pi,0)$ from 
Ref.\cite{norman1} (thin line) and our fits to the spectral function 
(thick line) at selected temperatures. The dashed line illustrates fit with 
$W=0$. Region of the fit is $|\omega|<80$meV.
b) Parameters of the fit as a function of $T$.
}
\label{fig1}
\end{figure}

The symmetrized experimental ARPES lineshapes for the  83K underdoped
BiSCCO sample of Ref.\cite{norman1} for the 
${\bf k}={\bf k}_F$ vector along the $(0,0)\to (0,\pi)$ 
direction at selected temperatures are shown in Fig.\ \ref{fig1}(a).
Below $T_c$ we wish to model these lineshapes by the spectral function 
corresponding to Eq.\ (\ref{g0}), which for  ${\bf k}={\bf k}_F$, 
$\epsilon_{{\bf k}_F}=0$ becomes
\begin{equation}
I_S(\omega)\propto A_0({\bf k}_F,\omega)={1\over \pi}
{\Gamma_1\over (\omega-\Delta_{\bf k}^2/\omega)^2+\Gamma_1^2}.
\label{spec}
\end{equation}
As noted in \cite{norman1}, this functional form indeed describes the data 
below $T_c$ well, 
after it is convolved in $\omega$ with a Gaussian of width $\sigma=13.5$meV 
representing the estimated experimental resolution. Qualitatively, 
$\Delta_{\bf k}$ sets the position of the quasiparticle peaks and $\Gamma_1$
(together with $\sigma$) sets their width. Note that for $\Gamma_1=\sigma=0$
the above lineshape consists of two 
$\delta$-functions at $\omega=\pm\Delta_{\bf k}$. 
It is thus clear that fairly large values of $\Gamma_1$ in Eq.\ (\ref{spec})
are needed in order obtain quasiparticle peaks of the correct width. 
Even for large $\Gamma_1$ the theoretical lineshape $I(\omega)$  tends to 
zero for $|\omega|$ larger then several $\Delta_{\bf k}$, while the 
experimental 
lineshape saturates to a finite value. Understanding this large-$\omega$
background presents a challenge for any theory of ARPES in the cuprates and
is beyond the scope of this paper. Here we focus on the low
energy region of the spectrum where we may  reasonably expect  the 
present simple model to be valid. 

For $\sigma\ll\Gamma_1$ one can easily estimate 
$I_S(0)\approx I_0(\Gamma_1\sigma^2/\Delta_{\bf k}^4)$ and 
$I_S(\Delta_{\bf k})\approx I_0\Gamma_1^{-1}$. The peak to valley ratio
\begin{equation}
\kappa\equiv
{I_S(\Delta_{\bf k})\over I_S(0)}\approx{\Delta_{\bf k}^4\over\Gamma_1^2
\sigma^2}
\label{pv}
\end{equation}
is independent of the 
unknown prefactor $I_0$ and can be easily extracted from the
raw data. Assuming fixed $\Delta_{\bf k}$ one can thus obtain reliable 
estimates 
of $\Gamma_1$ without performing detailed fits. In particular, application
of Eq.\ (\ref{pv}) to the data in Fig.\ \ref{fig1}(a) implies that $\Gamma_1$
grows by about a factor of 6 between $T=14$K and 100K.  
The above analysis also
implies that at low $\omega$ the lineshape depends crucially on the
experimental resolution $\sigma$. For instance decreasing $\sigma$ 
by a factor of 
2 the ratio $\kappa$ should grow by a factor of 4. Confirming this prediction
experimentally  would be a valuable test of the present model. 

Above $T_c$ Eq.\ (\ref{spec}) no longer provides a good fit for the data. 
The reason for this is the persistence (up to $\sim 200$K) 
of a well defined edge-like feature 
around $\omega=\Delta_{\bf k}$ along with a pronounced increase in the
low-$\omega$ density of states.  This behavior cannot be modeled by further 
increasing $\Gamma_1$ since the values needed to fix $I(0)$  would rapidly 
smear the edge. Inclusion of the phase fluctuations, i.e. finite $W$
in the averaged spectral function (\ref{avg}), rectifies this problem. 
Analytically
it is somewhat difficult to discuss the combined effect of $W$, $\Gamma_1$ and
$\sigma$
on the lineshape. Qualitatively one can show that the primary effect
of increasing $W$ is to ``fill in'' the gap. This is precisely what is needed
to describe the ARPES data above $T_c$.

Fig.\ \ref{fig1}(a) shows our fit to the symmetrized ARPES lineshapes for the 
underdoped sample
using a numerical computation of 
the full spectral function extracted from Eq.\ (\ref{avg}). Least square 
fits in which $\Delta_{\bf k}$, 
$\Gamma_1$ and $W$ were taken as free parameters,
were performed after convolving the spectral function 
$\bar A({\bf k}_F,\omega)$ with experimental resolution $\sigma=13.5$meV.
In the low energy region $|\omega|\lesssim 80$meV, where the simple model 
Green's function approach with $\omega$-independent scattering rate $\Gamma_1$
is expected to be valid, the fits are excellent for all temperatures. The
extracted parameters are displayed in Fig.\ \ref{fig1}(b). Both 
$\Delta_{\bf k}$
and $\Gamma_1$ behave in the way expected for an underdoped cuprate:
the gap is approximately constant across $T_c$ while the
scattering rate rises sharply below $T_c$ and saturates at higher 
temperatures. In the present model the large increase in $\Gamma_1$ is
required to wipe out the quasiparticle peaks.
$W$ also behaves as anticipated from the
above considerations. At low temperatures $W\simeq 0$ 
indicating that below $T_c$ the phase fluctuations are 
negligible; vortices appear only in tightly bound pairs and longitudinal 
fluctuations are suppressed by the 
Coulomb interaction. Above $T_c$ fluctuations become important (pairs unbind) 
and $W$ approaches the $T$-linear behavior consistent 
with (\ref{W1}). We stress here that a good fit to the data {\em requires}
$W>0$ above $T_c$. This is illustrated by the dashed line in Fig.\ 
\ref{fig1}(a)
which represents the fit to the 100K lineshape with $W=0$ and the gap value
restricted to $\Delta_{\bf k}\lesssim\Delta_{\bf k}(T=0)$\cite{remark11}. 
We note, however,
that in this model much of the observed low-$\omega$ density arises from the 
large value of $\Gamma_1$ in combination with the instrumental resolution.
It is possible (and is indeed suggested by the analysis of the STS data in the
next section) that the large $\Gamma_1$ contribution is an artifact, arising
from a combination of experimental resolution in the photoemission experiments
and inadequacies of our theoretical model. We therefore regard the values
of $W$ obtained here as underestimates. 

From the slope of $W(T)$, assuming that transverse 
fluctuations are dominant, we estimate $\alpha_v\simeq 0.009$ implying
the vortex 
core energy $E_c/V_0\simeq 360$. This value is much larger than the 
usual condensation energy in the vortex core $E_c\simeq 2V_0$ 
\cite{tinkham}. Within the Debye-H\"uckel approximation the vortex density
can be estimated as $\rho_v=\xi_c^{-2}(\pi/c_3)(1-\alpha_v)\simeq
 \xi_c^{-2}(T/T_{\rm KT})(V/c_3)$, for $\alpha_v\ll1$ and 
$k_BT_{\rm KT}\simeq V_0$. This implies, for the parameters extracted from 
ARPES data, $\rho_v\simeq 4.3\times 10^{-3}\xi_c^{-2}(T/T_{\rm KT})$. It is
remarkable that such a small density of vortices leads to significant 
broadening of the lineshape. We should also remark that for such a small 
density of vortices one may question the validity of Debye-H\"uckel 
approximation at temperatures in consideration. We emphasize, however, 
that only our interpretation of $W$ and in particular the estimates of 
$E_c$ and $\rho_v$ depend on the validity of this
approximation. Our analysis of the lineshapes is quite general, since we 
treat $W$ as a free parameter of the model.

We have performed similar fits for an  overdoped 82K BiSCCO sample
of Ref.\cite{norman1}. Our
results are consistent with $\Delta_{\bf k}$ vanishing close to $T_c$ and,
within statistical noise,  $W=0$ at all temperatures. This indicates that
phase fluctuations are unimportant in overdoped cuprates and the transition is 
essentially mean-field-like. 

We now consider
ARPES data at the Fermi crossing close to the gap node direction 
$(\pi,\pi)$. These indicate 
extended regions of gapless excitations above $T_c$ which grow in size 
with temperature\cite{ding1,norman2,norman1}. 
This is not reproduced by the simple model we have considered so far.
 The reason is that adding the $\hbar{\bf v}_s\cdot {\bf k}$
term to the quasiparticle energy [cf. Eq.(\ref{gq})] 
effectively depletes the local spectral function
for ${\bf v}_s$ parallel to ${\bf k}$ but enhances it by equal amount for 
opposite ${\bf v}_s$. Upon averaging over all directions of ${\bf v}_s$ 
(for fixed ${\bf k}$) 
the net effect is to broaden the mean-field lineshapes as seen in Fig.\ 
\ref{fig1}(a). Phase fluctuations cause no net depletion of the spectral 
weight near the gap nodes. 

In order to account for the ARPES data 
the form of $\Delta_{\bf k}$ must change.
Since the supercurrent flowing around individual 
vortices is pair-breaking, it is in principle possible that it will 
alter the  internal structure of the self-consistent gap function in 
addition to usual suppression of the order parameter in the core. 
A similar scenario has been proposed by Haas {\em et al.}\cite{haas}
who considered the effect of non-magnetic impurities on a $d$-wave
gap function. They found that it was possible to construct a gap function 
such that with increasing
disorder nodes indeed expanded into finite gapless arcs.
Intuitively this effect can be understood on the grounds that 
smaller gap near the nodes is more susceptible to the pair breaking. In the
present case pair breaking is caused by supercurrents rather than impurities,
but the physics remains the same. 

\begin{figure}[t]
\epsfxsize=8cm
\epsffile{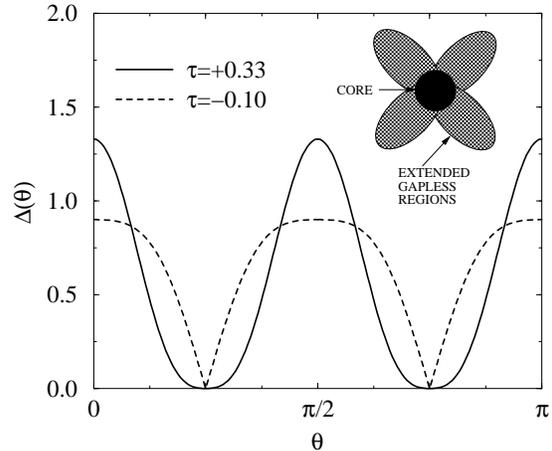}
\caption[]{Model gap function 
$\Delta(\theta)=\Delta_d\cos(2\theta)+\Delta'_d\cos(6\theta)$ with 
$\tau\equiv\Delta'_d/\Delta_d$ positive (extended nodes) and negative
(point nodes). Inset illustrates the proposed real space electronic 
structure of the vortex in a $d$-wave superconductor. 
}
\label{fig2}
\end{figure}

In order to substantiate this idea we have 
solved the self-consistent gap equation for a $d$-wave superconductor 
in the presence of {\em uniform} superflow. We considered a model gap
function of the form 
$\Delta_{\bf k}=\Delta_d\cos(2\theta)+\Delta'_d\cos(6\theta)$ with  the
appropriately generalized pairing interaction\cite{haas}. 
We found that, for a system
with $\Delta'_d<0$ in the absence of superflow, a transition to $\Delta'_d>0$  
occurs when sufficiently strong supercurrent flows in the direction close
to the nodal vector ${\bf k}_n$; {\em i.e.,} when
$|{\bf v}_s\cdot\hat{\bf k}_n|>b\Delta_d/\hbar k_F$, with $b\simeq 0.7$ a 
model dependent constant. The state with $\Delta'_d>0$ exhibits extended
gapless regions (cf. Fig.\ \ref{fig2}) while that with $\Delta'_d<0$ only the
usual point nodes. Extrapolating this behavior to the 
supercurrent flowing around the vortex we argue that a
region  of extended gapless excitations may form in the vicinity of the core. 
 Such a region would have 
the shape of a four leaf clover (schematically depicted in the inset
of Fig.\ \ref{fig2}) and a spatial extent of several $\xi_c$. A truly 
quantitative treatment of this effect is complicated because it involves 
self-consistently solving the $d$-wave vortex problem which is a highly
non-trivial task\cite{franz1}. We note, however,  that recent STS data
on vortices in BiSCCO\cite{renner3}, show a peculiar pseudogap behavior
near the vortex core, which may be indicative of a formation of the gapless
regions around a vortex proposed above. 

Assuming that this picture is correct, it is clear that in the  
vortex-antivortex plasma above $T_c$, upon averaging over fluctuations, ARPES
will detect a gap function strongly suppressed for ${\bf k}$
close to the nodes. Furthermore, 
with increasing temperature the volume fraction of gapless regions will grow
(since the vortex density grows) leading to larger gapless areas on the
Fermi surface in agreement with experimental observation.

\subsection{Tunneling}

\begin{figure}[t]
\epsfxsize=9.5cm
\epsffile{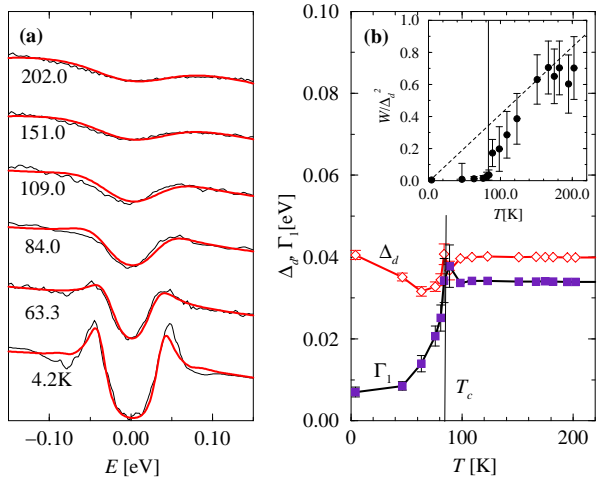}
\caption[]{a) Fit to the STS data of Ref. \cite{renner1} at selected 
temperatures. b) Parameters of the fit as a function of $T$. Above $T_c$
both $\Delta_d$ and $\Gamma_1$ are fixed to their values at 84K. 
}
\label{fig3}
\end{figure}

Tunneling conductance, the quantity measured by STS, is related to the
spectral function of the superconductor by 
\begin{equation}
g(E)=\int_{-\infty}^{\infty}d\omega f'(\omega-E)\sum_{\bf k} 
|M_{\bf k}(\omega)|^2\bar A({\bf k},\omega),
\label{cond}
\end{equation}
where $f$ is the Fermi function and $M_{\bf k}(\omega)$ is the tunneling
matrix element, usually approximated by a constant. Tunneling conductance
reflects the spectral function averaged over the entire Brillouin zone and
broadened in the energy variable by the Fermi function. Thus, unlike in the 
ARPES lineshape function $I_S(E)$, quasiparticle
peaks in $g(E)$ will be broadened even in the absence of scattering and at 
$T=0$. For measurements performed on similar samples
one would thus expect ARPES spectra to be much sharper than STS spectra, at any
temperature. Fig.\ \ref{fig3}(a) displays $g(E)$ as measured by STS on 83K 
underdoped BiSSCO sample of Ref. \cite{renner1}.
Surprisingly, we observe that at lowest 
available temperatures STS line is in fact {\em sharper} than the ARPES line. 
As discussed in the previous section the 
broadening of the ARPES line comes exclusively from the scattering and 
experimental resolution. Therefore, inspection of the raw data
suggests that the scattering rate $\Gamma_1^{\rm STS}$ will be much smaller 
than $\Gamma_1^{\rm ARPES}$. This conclusion is indeed borne out by the 
detailed fitting procedure carried out below. This is a rather surprising 
result which we discuss more fully in the next section. 

As seen from Eq. (\ref{cond}), modeling of the
tunneling conductance requires knowledge of the band structure away from the 
Fermi surface and is therefore somewhat more involved than that of ARPES. 
Nevertheless, we 
find that assuming a simple free electron dispersion with cylindrical
Fermi surface and $\Delta_{\bf k}=\Delta_d\cos(2\theta)$, provides a 
reasonable fit for the data  at 
low temperatures, provided that one compensates for the asymmetric 
background conductance and assumes $\theta$-dependent matrix element
$M_{\bf k}(\omega)\propto |\cos(2\theta)|$. The latter assumption 
is motivated by 
band structure calculations\cite{band1} which indicate that tunneling between
layers is dominated by the regions on the Fermi surface close to the $(0,\pi)$
points. As shown in Fig.\ \ref{fig3}(a), this $\theta$-dependence allows to 
simultaneously account for the sharp quasiparticle peak and the wide gap
in the STS line. Assuming $M_{\bf k}(\omega)$=const leads to broader peaks and
sharper V-shaped gap structure, inconsistent with experimental data. We
further remark that our simple model does not (and is not expected to)
explain the pronounced
dip feature appearing in the spectrum at higher energies, whose origin is at 
present unknown.
Over the range $T<T_c$ $\Gamma_1^{\rm STS}$  has similar 
qualitative behavior as $\Gamma_1^{\rm ARPES}$
[compare Figs. \ref{fig1}(b) and \ref{fig3}(b)], but remains
about one order of magnitude smaller. We have attempted to reconcile the 
two sets of data by considering a more realistic band structure and 
$\theta$-dependent scattering rate in 
Eq.(\ref{cond}). However, even under the most
favorable conditions $\Gamma_1^{\rm STS}$ remains factor of 5-6 smaller than 
$\Gamma_1^{\rm ARPES}$. 

Above $T_c$ the STS line does not have enough features to permit a 
meaningful three parameter fit. In particular the sharp gap edge completely 
disappears above 100K which leads to ambiguity in defining $\Delta_d$ from 
the data.  Based on our previous finding (from the ARPES data) that 
both $\Delta_d$ and $\Gamma_1$ exhibit only weak $T$-dependence above $T_c$,
we fix these two parameters at their respective values at 84K (40meV and 
34meV) and extract the temperature dependence of $W$. This appears to us as a 
reasonable procedure  which we further check by performing two-parameter 
fits with $\Delta_d$ held constant or slowly decreasing with temperature
as implied by Fig.\ \ref{fig1}(b). $W$ is shown in the inset
to Fig.\ \ref{fig3}(b).  Its qualitative behavior is similar to that of 
$W^{\rm ARPES}$ but  the amplitude is roughly factor of 
10 {\em larger}. The difference is caused in part by the much smaller 
value of $\Gamma_1$ discussed above. Part of the discrepancy between 
$W^{\rm ARPES}$ and $W^{\rm STS}$ can 
presumably be attributed to differences 
in material and experimental uncertainties, as well as the failure of our
fit to account for the temperature variation of $\Gamma_1$, which according
to Fig. \ref{fig1}(b) grows by another factor of 2 between 80K and 200K.  
Nevertheless, after accounting for these factors, considerable discrepancy
remains in place which is not understood at present.
We estimate $\alpha_v\simeq 0.097$ which implies the
vortex core energy $E_c\simeq 22V_0$ and vortex density 
$\rho_v\simeq 8.7 \times 10^{-2}\xi_c^{-2}(T/T_{\rm KT})$. 
The value of $E_c/V_0$ is still large compared to 
the conventional estimate of the condensation energy in the core
$E_c\simeq 2V_0$ \cite{tinkham}, 
but is consistent with large core energy  $E_c\simeq 26V_0$
deduced from lower critical field measurements of YBa$_2$Cu$_3$O$_{6.95}$ at
$T=0$\cite{liang}. We note that $E_c$ is a 
cutoff-dependent quantity and therefore the precise numerical value quoted
here should be accepted with that in mind. It is also possible that the 
large value of the ratio $E_c/V_0$ is due to an unusually small $V_0$ rather 
than unusually large $E_c$. Indeed, the vortex core energy is typically
of the order of Fermi energy. Estimate of $V_0$ given in Ref.\ \cite{millis1}, 
$5$meV$<V_0<10$meV, implies $E_c\sim 0.1-0.2$eV, in reasonable agreement with 
many theories of underdoped cuprates which suggest a Fermi energy of the order
of the exchange constant $J\sim 0.15$eV.

\section{Discussion}

The qualitative behavior of ARPES and STS lineshapes  
in underdoped BiSCCO clearly establishes the existence of a 
scattering mechanism which becomes operative at $T>T_c$ and 
which acts primarily to fill in the gap at low energies. We have shown
that transverse phase fluctuations associated with proliferation of
unbound vortex-antivortex pairs in the system provide a reasonable explanation
for this scattering. Our analysis also indicates that longitudinal (spin
wave) fluctuations are almost completely suppressed, above and below $T_c$.
It has been proposed \cite{roddick1,emery2} that in high-$T_c$ materials,
longitudinal phase fluctuations governed by the XY Hamiltonian (\ref{ham})
are important in that they significantly contribute to the observed temperature
dependence of the magnetic penetration depth\cite{bonn}. We have calculated 
the broadening of the spectral function which would be caused by these 
fluctuations, and found it to be much greater than the experimental data 
would permit. We therefore conclude that longitudinal fluctuations are
suppressed, perhaps by the Coulomb interaction as suggested in Ref.\ 
\cite{millis1}.

Quantitatively there exists considerable discrepancy between
the parameters describing the ARPES and STS lineshapes, in particular 
the single particle 
scattering rate $\Gamma_1$ and phase fluctuation broadening $W$.
Since the discrepancy is apparent at low temperatures and in overdoped 
cuprates we are led to believe that the problem lies primarily 
in our lack of detailed understanding of the lineshapes
rather than the physics of phase fluctuations above $T_c$.
The most disturbing is almost an order of magnitude difference between
$\Gamma_1^{\rm ARPES}$ and $\Gamma_1^{\rm STS}$ found below $T_c$, which
is implied directly by the raw data.  In view of the fact that both 
measurements pertain to underdoped BiSCCO crystals with similar 
critical temperatures, it appears unreasonable to attribute such a large 
discrepancy to the material differences. We speculate that the large scattering
rate needed to fit the ARPES data is an artifact related to our incomplete
understanding of the photoemission process in the superconductor
which is theoretically not completely understood even in simple 
metals\cite{cardona}. Tunneling spectroscopy, on the other hand, is a 
technique well 
established in superconductors. We therefore surmise that parameters 
obtained from STS more directly reflect the underlying physics. Indeed 
$\Gamma_1^{\rm STS}\simeq 8$meV at 4.2K is comparable to the scattering rates
deduced from transport measurements\cite{basov,ore} on underdoped cuprates, and
$E_c^{\rm STS}\simeq 22V_0$, although large for a conventional superconductor,
is perhaps not unreasonable in cuprates\cite{liang}.
Consequently, ARPES 
lineshapes appear to reflect significant extrinsic broadening of unknown
origin. The puzzling aspect of this interpretation is that
the additional physics in the ARPES spectra 
enters as a {\em multiplicative} rather than 
additive factor to the apparent scattering rate; cf. 
$\Gamma_1^{\rm ARPES}\simeq 8\Gamma_1^{\rm STS}$ over the entire 
temperature range below $T_c$, in which $\Gamma_1$ changes by a factor of 6.
It is also possible that in the cuprates the $c$-axis tunneling matrix
element $M_{\bf k}(\omega)$ introduced in Eq.\ (\ref{cond}) is itself 
anomalous. Improving the energy resolution $\sigma$ of ARPES could
shed some light on this issue. As noted below Eq.\ (\ref{pv}) the ARPES 
lineshapes are strongly affected by experimental resolution at small $\omega$.
If the model Green's function (\ref{g0}) is correct, a factor of two
improvement in $\sigma$ 
should lead to considerable decrease in the measured intensity at
$\omega=0$ but almost no change in the width or height of the quasiparticle 
peaks at $|\omega|=\Delta_{\bf k}$.

Finally 
we note that sizable transverse phase fluctuations implied by this work 
will also affect other 
properties of the underdoped systems, such as the electronic specific heat, 
fluctuation diamagnetism and transport. Vortices existing above $T_c$ should 
also generate local magnetic fields which are zero on average but have a
non vanishing variance. If such fields could be detected, e.g.  by
muon spin rotation experiment, this would constitute a direct evidence for 
the phase fluctuation model of the pseudogap phase.

\acknowledgments
The authors are indebted to J. C. Campuzano and Ch. Renner for providing their 
experimental data and to M. R. Norman, S. Teitel and Z. Te\v{s}anovi\'{c} for 
insightful discussions. This work was supported by NSF
grants DMR-9415549 (M.F.) and DMR-9707701 (A.J.M.) and by the Theoretical
Interdisciplinary Physics and Astronomy Center at the Johns Hopkins University.

\appendix
\section*{Gorkov equations in the presence of superflow}
 
Real-space Gorkov equations\cite{agd} generalized to 
anisotropic superconductors read
\begin{eqnarray}
(\omega-\hat{\cal H}_e){\cal G}({\bf r}_1,{\bf r}_2;\omega)+\hat\Delta
{\cal F}^+({\bf r}_1,{\bf r}_2;\omega) 
&=&\delta({\bf r}_1-{\bf r}_2),\nonumber \\
(\omega+\hat{\cal H}_e^*){\cal F}^+({\bf r}_1,{\bf r}_2;\omega)+
\hat\Delta^*{\cal G}({\bf r}_1,{\bf r}_2;\omega) &=&0.\nonumber\\
\label{gor1}
\end{eqnarray}
Here $\hat{\cal H}_e=[i\nabla-(e/c){\bf A}]^2/2m-\epsilon_F$ is the single 
electron Hamiltonian and $\hat\Delta$ is the gap operator for spin singlet
superconductivity defined as 
\begin{equation}
\hat\Delta{\cal F}^+({\bf r}_1,{\bf r}_2;\omega)=\int d^2r' \Delta({\bf r}_1,
{\bf r}')
{\cal F}^+({\bf r}',{\bf r}_2;\omega).
\label{gap0}
\end{equation}
$\Delta({\bf r}_1,{\bf r}_2)$ is the gap function which is
in general a nontrivial  function of both electron coordinates in the 
anisotropic superconductor. We are interested in the state of uniform superflow
induced by the gap function of the form
\begin{equation}
\Delta({\bf r}_1,{\bf r}_2)=\Delta_0({\bf r}_1-{\bf r}_2)e^{i{\bf q}
\cdot({\bf r}_1+{\bf r}_2)}
\label{gap1}
\end{equation}
and ${\bf A}=0$. The easiest way to solve (\ref{gor1}) 
for ${\cal G}$ is to perform 
a gauge transformation to the gauge where the order parameter is real and
independent of the center of mass coordinate ${\bf R}=({\bf r}_1+{\bf r}_2)/2$:
\begin{eqnarray}
\Delta({\bf r}_1,{\bf r}_2)&\to&\Delta({\bf r}_1,{\bf r}_2) e^{-i{\bf q}
\cdot({\bf r}_1+{\bf r}_2)},
\nonumber\\ 
{\bf A}&\to&{\bf A}-{c\over e}{\bf q},
\nonumber\\
{\cal G}({\bf r}_1,{\bf r}_2;\omega)&\to& {\cal G}({\bf r}_1,{\bf r}_2;\omega)
e^{-i{\bf q}\cdot({\bf r}_1-{\bf r}_2)},\nonumber \\
{\cal F}^+({\bf r}_1,{\bf r}_2;\omega)&\to& {\cal F}^+({\bf r}_1,{\bf r}_2;
\omega)
e^{i{\bf q}\cdot({\bf r}_1+{\bf r}_2)}.
\label{gauge}
\end{eqnarray}
It is easy to verify that under such transformation Eqs.\ (\ref{gor1})
remain invariant\cite{agd}. In the new gauge Gorkov equations are manifestly
translationaly invariant, i.e. independ of ${\bf R}$.
Fourier transforming in the relative coordinate 
${\bf r}={\bf r}_1-{\bf r}_2$ leads to algebraic equations for ${\cal G}$ and
${\cal F}^+$  of the form 
\begin{eqnarray}
(\omega-\epsilon_{{\bf k}-{\bf q}}){\cal G}({\bf k},\omega)+\Delta_{\bf k}
{\cal F}^+({\bf k},\omega) &=&1\nonumber\\
(\omega+\epsilon_{{\bf k}+{\bf q}}){\cal F}^+({\bf k},\omega)+ 
\Delta_{\bf k}{\cal G}({\bf k},\omega) &=&0,
\label{gor2}
\end{eqnarray}
where $\epsilon_{\bf k}={\bf k}^2/2m-\epsilon_F$. The solution for ${\cal G}$
is 
\begin{equation}
{\cal G}_{\bf q}^{-1}({\bf k},\omega)=\omega-\epsilon_{{\bf k}+{\bf q}}-
{\Delta_{\bf k}^2\over \omega+\epsilon_{{\bf k}-{\bf q}}}.
\label{gqq1}
\end{equation}
Expanding $\epsilon_{{\bf k}\pm{\bf q}}$ to leading order in 
${\bf q}$ we obtain
\begin{eqnarray}
{\cal G}_{\bf q}^{-1}({\bf k},\omega)&=&(\omega-\eta)-\epsilon_{{\bf k}}-
{\Delta_{\bf k}^2\over (\omega-\eta)+\epsilon_{{\bf k}}}\nonumber\\
&=&{\cal G}_0^{-1}({\bf k},\omega-\eta),
\label{gqq2}
\end{eqnarray}
with $\eta\equiv{\bf v}_F({\bf k})\cdot{\bf q}\simeq{\bf k}\cdot{\bf v}_s$. 
As a final step
we transform ${\cal G}$ back to the original gauge with ${\bf A}=0$.
According to  (\ref{gauge}) this amounts to simply replacing 
${\bf k}\to{\bf k}-{\bf q}$ on the right hand side of Eq.\ (\ref{gqq2}). 
We thus obtain the desired expression (\ref{gq}). In deriving this result 
we have assumed for simplicity a free particle form of the single electron 
Hamiltonian $\hat{\cal H}_e$. Evidently, the calculation remains valid for 
more complicated Hamiltonians.


\begin{references}
\bibitem{note4} We stress that the present model can be equally well
formulated without defining a superconducting mass $m$. However,
such formulation leads to somewhat awkward notation and we therefore choose
to follow the standard literature in defining the superfluid velocity and 
other quantities.
\end{references}
\end{document}